# Laser-Induced Gas-Phase Transfer and Direct Stamping of Nanomaterials: Comparison of Nanosecond and Femtosecond Pulses

Nathan T. Goodfriend[1], Inam Mirza[1], Alexander V. Bulgakov[1], Eleanor E.B. Campbell[2,3], Nadezhda M. Bulgakova[1,*]

1. FZU – Institute of Physics of the Czech Academy of Sciences, Na Slovance 1999/2, 182 00 Prague, Czech Republic
2. EastCHEM, School of Chemistry, University of Edinburgh, David Brewster Road, Edinburgh EH9 3FJ, United Kingdom
3. Department of Physics, Ewha Womans University, Seoul 03760, Republic of Korea

[*]Corresponding author: bulgakova@fzu.cz

## Abstract:

The two-dimensional nanomaterial, hexagonal boron nitride (hBN) was cleanly transferred via a blister-based laser-induced forward-transfer method. The transfer was performed utilizing femtosecond and nanosecond laser pulses for separation distances of ~16 and ~200 μm between a titanium donor film deposited on a glass substrate and a silicon/silicon dioxide receiver. Transfer efficiency was examined for isolated laser pulses as well as for series of overlapping pulses and single layer transfer was confirmed. It was found that hBN is transferable for all tested combinations of pulse duration and transfer distances. The results indicate that transfer proceeds via direct stamping for short donor-to-receiver distances while, for the larger distance, the material is ejected from the donor and lands on the receiver. Furthermore, with overlapping pulses, nanosecond laser pulses enable a successful printing of hBN lines while, for fs laser pulses, the Ti film can be locally disrupted by multiple pulses and molten titanium may be transferred along with the hBN flakes. For reproducibility, and to avoid contamination with metal deposits, low laser fluence transfer with ns pulses and transfer distances smaller than the blister height provide the most favourable and reproducible condition.

## Introduction

Two-dimensional (2D) materials are attracting increased attention due to their novel physical properties, which appear at reduced planar dimensions. The reduced dimensionality restricts the movement of charge and hence information transfer by specific pathways across the material [1]. Graphene as the first discovered 2D material has become a flagship material in the field due to its exceptional conductive properties [2]. Demonstration of the unique properties of graphene created a surge of interest in seeking new 2D materials and exploring their unique features for various applications [3-6]. Transition metal dichalcogenides (TMDs) fill a specific niche within the developing nanoelectronic framework as they exhibit a variety of bandgaps [7-12]. Hexagonal boron nitride (hBN) is also of key interest as a 2D insulator and is often used to encapsulate other semiconducting or semimetallic 2D materials [13-17]. In this work, we utilize exfoliated hBN as a case material to demonstrate an effective and clean method of transferring 2D materials via a blister-based laser methodology.

There are multiple methods of synthesizing 2D materials and their placement on desired substrates for applications [5,12,15,18-23]. However, the latter is still limited with regard to scalability, cleanliness, and/or precision [19, 24,25]. One of the common methods is exfoliation, which often produces a dispersion of nanomaterials that are difficult to position precisely [26]. Also, the exfoliation method (micromechanical or using scotch tape) gives a low yield [27]. Liquid exfoliation is another method for 2D material generation, which results in a relatively narrow distribution of material thickness [28].

However, when dried from solution, the material flakes are randomly distributed over the surface and their assembly into a desired structure is complicated.

Other methods that produce a more controlled set of 2D nanomaterials are chemical vapor deposition (CVD) and molecular beam epitaxy (MBE). These methods can generate controlled layer numbers and large-scale growth on metal or ceramic substrates [29,30-32]. High temperatures and chemical atmospheres can be used to limit the growth to desired locations [33-35]. To assemble the nanomaterials into workable devices, a multistep polymer-based pick-and-place method is often used [18,21]. This polymer-based method results in impurities, microbubble formation, and ripples and it can be challenging to transfer individual material pieces to construct complex devices [36]. The method discussed in this paper is a variation of the laser-induced forward transfer (LIFT) method which is successfully used for printing various nanomaterials for applications in micro/nanoelectronics, sensing, and biomedicine [25,37-41]. It is called "blister-based" LIFT (BB-LIFT) and its uniqueness lies in the possibility of avoiding the direct laser heating of the transferred nanomaterial, thus minimizing any thermal damage [42-45]. A variation of BB-LIFT, blister-actuated LIFT, is used for printing liquid droplets with a high lateral resolution [46,47]. The BB-LIFT method can efficiently supplement other existing transfer methodologies due to its physical transfer mechanism enabling contamination-free transfer as opposed to chemical-based methods.

The BB-LIFT method is potentially suitable for the transfer of fragile 2D materials [45,48]. Blister-based LIFT utilizes a pulse of energy, which is low enough to only modify a metallic donor film at its interface with a transparent substrate [44-46,48.49]. The pulsed laser irradiation causes a blister-like deformation of the film which serves as a dynamic release layer (DRL) [49-51], thus physically ejecting nanomaterial deposited on the top of the metal film with a high degree of directionality [43-45]. In previous work by Goodfriend et al. [45], it has been shown that the BB-LIFT technique can safely transfer low-dimensional materials, such as $MoSe_2$ and $MoS_2$, by femto- and nanosecond laser pulses under vacuum conditions to distances on the order of millimeters when using 200-300 nm thick titanium films as DRLs. Recently, the BB-LIFT transfer of CVD-grown graphene was successfully demonstrated [22] while it was shown that using LIFT without a DRL also allows direct printing of pixels from aqueous graphene oxide films with simultaneous oxide reduction [52]. Further experiments in laser-induced backward-transfer (LIBT) have demonstrated monolayer flat graphene transfer [38]. In [22], a thin (50 nm) nickel film was used as the DRL while crumpled graphene transfer was also achieved with thicker, 420-1900 nm, aluminum DRLs [48]. This work aims at exploring and extending the BB-LIFT method for transferring the insulating 2D material, hBN. Comparisons are made for transfer with nano- and femtosecond laser pulses as well as for short and longer distances between donor and receiver substrates.

The laser pulse duration has a strong impact on the blister formation in the BB-LIFT geometry. There are two mechanisms that may dominate for different irradiation conditions [50,53]. At relatively low laser fluences or longer (nanosecond) pulse durations, the metal film is rapidly heated and expands forming a blister due to thermal stress [50,54]. When the film cools again, it can be re-contracted. For shorter pulse durations, on the order of 100 fs, the metal at the interface with the transparent substrate can be rapidly ablated in a confined fashion. The ablated material, being at high pressure, expands, causing a blister to form via stretching of the film [50,53]. Both mechanisms should lead to the ejection of nanomaterials located on the top of the DRL in a highly directed manner and, in both cases, the nanomaterials typically survive in the transfer process, receiving negligible to zero heat or irradiation damage [44,45,50,55]. However, there are considerable variations in the velocity of the ejected material depending on the irradiation conditions [43,45,49,56-58]. The general tendency appears to be as follows. The higher the laser fluence and the thinner the film, the higher the velocity of the transferred nanomaterial.

In this paper, the mechanisms of BB LIFT are studied using femtosecond and nanosecond laser pulses to transfer 2D materials. The investigations have been performed under ambient atmospheric conditions. The results are analyzed via polarized light microscopy, atomic force microscopy (AFM), and Raman spectrometry. We demonstrate the efficacy of transfer via the BB-LIFT methodology and discuss the underlying mechanics with the goal of determining the optimum conditions for a controllable assembly of 2D nanomaterials.

## Methodology

Four different transfer conditions are investigated in this paper, as schematically illustrated in figure 1. These conditions investigate the effect of the distance between the receiver and donor substrates and how the laser pulse duration influences the transfer process. The donor substrate consists of a 220-nm thick titanium film (DRL) deposited on a glass substrate. The hBN flakes were placed on the DRL by using the "scotch tape" method. A small amount of boron nitride was removed from a commercial bulk sample and placed on a long stretch, approximately 50 cm, of sellotape. The sellotape is then folded onto itself and peeled away, splitting the hBN. This is done repeatedly, approximately 15 times, ensuring to not utilize the same location of the sellotape. This results in sellotape coated by hBN flakes of a variety of thicknesses ranging from monolayer to hundreds of layers. The sellotape is then smoothly pressed on the titanium DRL, ensuring that no air gaps or bubbles are present. The sellotape was removed leaving many fragments of 2D nanomaterial on the donor substrate along with some glue residues from the sellotape. The glue was removed prior to carrying out the BB LIFT transfer experiments by immersing the donor substrate in a shallow bath with dichloromethane until no glue could be visibly seen under a microscope. This method, although not the most advanced way to deposit 2D nanomaterials, is a simple and practical way to provide substrates suitable for the BB LIFT experiments. The majority of deposited hBN flakes are multilayered although single-layer flakes are also present.

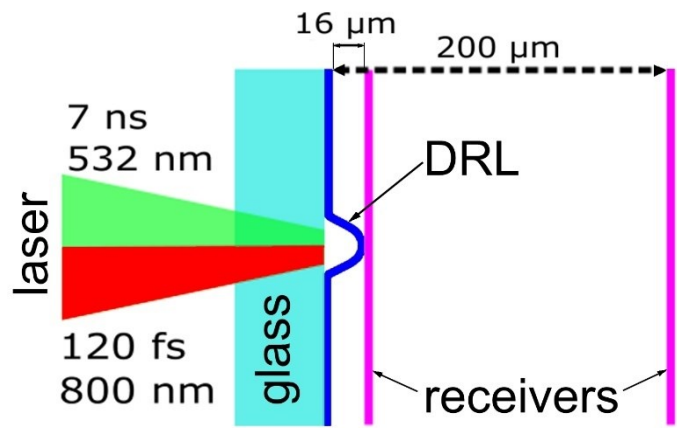

*Figure 1: Schematics of BB-LIFT in which two laser irradiation regimes and two donor-to-receiver distances were used.*

Two lasers were used in the experiments, a Ti-sapphire laser (800 nm wavelength, 120 fs pulse duration, Gaussian spatial profile with 1/e radius of 140±4 μm on the DRL location) and a nanosecond Gaussian pulse (532 nm, 7 ns) with 1/e radius of 155 μm ±10.

The donor and receiver were clamped together using rubberized low-pressure clamps. A controlled separation distance was provided by using either two stacked spacers of aluminum foil ~8 μm thick or glass coverslips of ~200 μm. The donor-receiver arrangement was mounted on a tilt-adjustable *xyz* translation stage. It should be noted that there is an uncertainty in the donor-to-receiver distance on the order of 1 μm. The smaller gap is referred to below as the near-contact regime.

The peak laser pulse fluence $F_0$ was varied within the ranges already known to produce blisters on Ti films of a similar thickness [42,44,49,50]. For ns-produced blisters, the corresponding $F_0$ values were varied from 135 ±15 mJ/cm², at which smooth non-cracked blisters were formed, to 290±30 mJ/cm² per pulse where the blister could crack and occasionally burst open. For fs-produced blisters, the fluences were in the range from 180 ±10 mJ/cm² to 260 ±10 mJ/cm² per pulse. At 220-260 mJ/cm², the blisters were liable to burst with the actual threshold value for bursting depending on small variations in the film thickness. This film disruption can also result from a large difference in the generated stresses which, for fs irradiation, can reach a Gigapascal level [58] while, for ns irradiation, it substantially dissipates already during irradiation [57], thus providing a gentler blistering.

The blisters transfer the deposited 2D materials onto a silicon surface with a 270 nm thick oxide layer that increases the optical contrast and identification of the hBN under an optical microscope [59,60]. Two sets of experiments were performed for both femtosecond and nanosecond pulses. In one set, individual spots were irradiated. In another set, the irradiation regions were overlapped by moving the position of the laser beam by 100 µm after each shot, thus creating a line of very close proximity blisters. To characterize the initial and transferred 2D material, optical microscopy (Olympus BX43), atomic force microscopy (AFM) and Raman spectroscopy (XploRA Nano, Horiba Scientific) were used.

## Evaluation of blister heights

The height of laser-induced blisters can be estimated using a simplified geometrical model in application to both interface ablation [42] and thermal expansion [57] mechanisms of blistering. According to the model, the blister height as a function of the temperature and the radius of the heated area is

$$h \approx r \frac{1-\cos\sqrt{6K_e \Delta T}}{\sin\sqrt{6K_e \Delta T}} \quad (1)$$

where $h$ is the blister height, $r$ is the radius of the irradiation spot, $K_e$ is the coefficient of linear expansion, and $\Delta T$ is the temperature rise upon laser heating. The $\Delta T$ value can be estimated as [57]

$$\Delta T = \frac{(1-R)F_0}{c_p d} \quad (2)$$

Here, $R$ is the reflectivity, $c_p$ is the specific heat of the material and $d$ is the thickness of the DRL. Note that here $\Delta T$ is an average temperature rise across the heat-affected zone related to the average laser fluence.

As noted in [57], the bump height evaluated by Eqs. (1)-(2), based on equilibrium assumptions, can be underestimated as it neglects dynamic effects. The swiftly expanding film can be overstretched compared to the equilibrium situation and the final bump height can be larger. Figure 2 shows the predicted minimum height of a laser-induced blister. The conditions for the present study, as evaluated using Eqs. (1)-(2), are outlined by green and red lines respectively for nano- and femtosecond laser pulses with their projections on the temperature-height plane. For estimations, the specific heat values were used at the elevated temperatures [61] to which the Ti film was heated. For this, an iterative procedure was used to fit $c_p(T)$ and $T$ values for each laser fluence. The reflectivities of a Ti film deposited on a glass substrate were calculated for 532 nm and 800 nm using the Filmetrics® website [62] for the system "air – 1 mm SiO₂ – Ti" which yielded respectively 0.392 and 0.449. Figure 2 illustrates that both laser fluence and irradiation spot radius affect the height of the blister expansion. The maximum fluences used in our irradiation regimes at which smooth non-cracked blisters were formed, lead either to heating the film to the melting point for ns laser pulses or close to the melting point for fs pulses, assuming thermal equilibrium across the film.

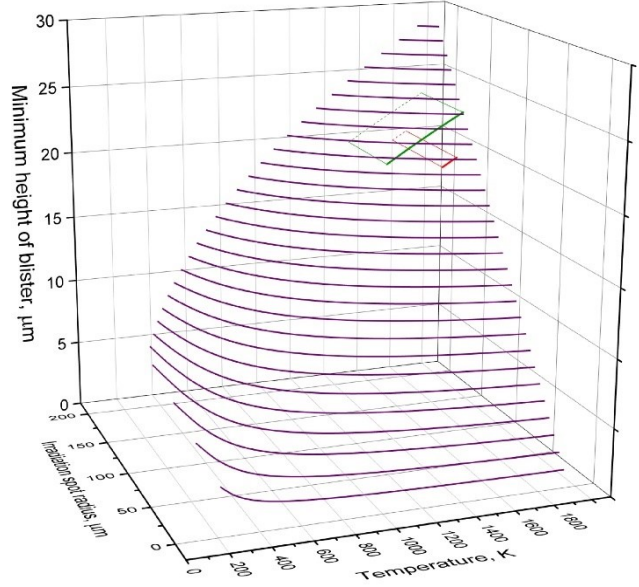

*Figure 2: Minimum height of the laser-induced blister as a function of the temperature and the radius of the irradiated spot for titanium film 220 nm thick, evaluated by Eqs. (1)-(2). Contour lines are given with the step Δh = 1 μm. The green and red lines show temperature-height ranges for laser fluences used in the study in nano- and femtosecond regimes respectively. For convenience, their projections to the temperature-height plane are added (dotted lines).*

In the ns irradiation regime, the temperature, evaluated as $T = T_0 + \Delta T$, is ~1540 K for the minimum used laser fluence of 135 mJ/cm² while the maximum temperature (~2300 K at $F_0 = 290$ mJ/cm²) exceeds the melting point, $T_{melt} = 1943$ K. However, complete melting is not reached as can be found from the estimation of the melting fraction $f$ as $f = c_p(T - T_{melt})/\Delta H_{melt}$ [63] with $\Delta H_{melt} = 391$ J/g to be the heat of fusion [64]. Using the heat capacity values at $T_{melt}$ for solid and liquid phases [61], $c_p^s(T_{melt})$ and $c_p^l(T_{melt})$ respectively, and applying the iterative procedure with the expression for the partially molten material $c_p^{s-l}(T_{melt}) = (1-f)c_p^s(T_{melt}) + fc_p^l(T_{melt})$ [65], we estimate $f \approx 78\%$. Thus, the green line in figure 2 starts from 1240 K and ends at $T = T_{melt}$. Further increasing the fluence can potentially result in a larger blister height but it can also cause a rupture of the DRL, particularly due to complete melting, leading to damage to the transferred material, as we observe experimentally for the highest investigated fluences. It should also be mentioned that, in ns irradiation regimes, a part of the absorbed energy will be transferred to the substrate already during the laser pulse and, hence, the melting fraction of the Ti film [57] can be lower than estimated above.

For the fs irradiation regime, the smooth non-cracked blisters were formed in a narrow range of laser fluences. As a result, the temperature evaluated for the minimum fluence of $F_0 = 180$ mJ/cm², assuming thermal equilibrium, is ~1800 K and, for the upper limit of the smooth blister formation, $F_0 = 220$ mJ/cm², it is ~1880 K (figure 2, red line), close to melting but still below the melting point. According to the evaluations made above, for the Gaussian beam shape in the fs irradiation regime within the studied fluence range, complete melting across the film cannot be achieved. However, the mechanisms of the film heating and the blister formation with ultrashort laser pulses are more complicated and can include the vaporization of titanium at the glass-DRL interface [50,53] before the temperature is equilibrated across the film, in contrast to the ns case. We note that the melting threshold of titanium irradiated with femtosecond laser pulses at 800 nm wavelength is around 0.2 J/cm² (peak fluence) [66]. Due to very rapid electron-lattice thermalization in titanium [67], a thin layer of order of the light penetration depth at the interface with the transparent substrate can experience melting and even ablation/disintegration in several picoseconds at fluences above 0.2 J/cm² while the heat propagation across the film may take tens of picoseconds. The ablation of titanium at the film-substrate interface, if it happens, will consume a part of the absorbed laser energy and, thus, decrease the film temperature which can reduce film heating throughout its thickness, even at the highest peak fluences studied here.

A high temperature gradient across the film may contribute to the disruption of the film mentioned above for the fluence range of 220-260 J/cm$^2$.

# Results and Discussion

All the blisters created in this work by nanosecond and femtosecond laser irradiation, transfer the hBN flakes from localized regions whose size is approximately equal to the size of the blister as illustrated in figure 3. The BB-LIFT technique does not discriminate in transfer between monolayer and multilayer flakes. The majority and, in some cases, all flakes are removed from the blister areas on the donor substrate regardless of the flake thickness. This can be seen in selected images in figure 3 as the hBN flakes show clear color variation depending on the number of layers/flake thickness.

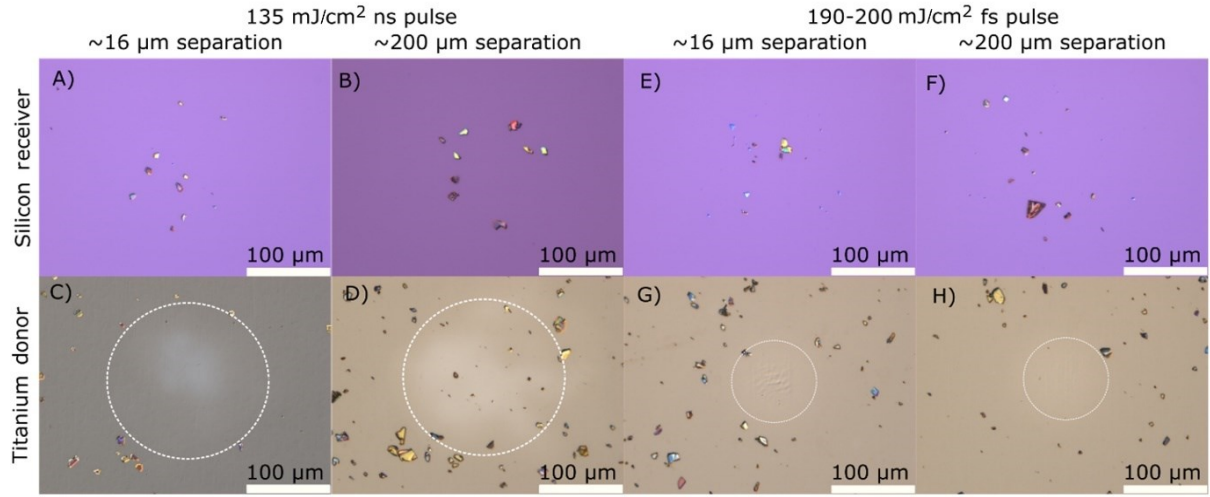

*Figure 3: Optical microscope images of the transferred hBN flakes on a silicon substrate (upper row) and the corresponding blister areas from which the hBN flakes were transferred. On the left, ns BB-LIFT is shown at a laser fluence of 135±15 mJ/cm$^2$ for short (A,C) and long (B,D) donor-to-receiver distances. On the right, fs transfer cases are presented for the short donor-to-receiver distance (E,G) at $F_0$ = 190 ± 10 mJ/cm$^2$ and the long distance (F,H) at 200 ± 10 mJcm$^{-2}$. The white dotted circles outline the approximate blister expansion regions, generally smaller than the laser spot size*

The amount of hBN flakes transferred is largely controlled by the size of the blister, the laser fluence, and the density of material deposited on the donor substrate. However, it can be seen from figure 3 that, for ns laser-induced blisters, the distance between the donor and receiver substrates plays an important role. The ns blister areas have almost no hBN left after irradiation when the donor-to-receiver distance is ~16 μm while, for the distance of ~200 μm, some flakes remain untransferred. In the case of fs laser pulses, there is little distinction between the remaining deposit on the blisters produced for both distances. According to Eq. (1), the minimum height of the blister formed due to purely thermal effects for ns irradiation should be greater than ~18 μm in our experiments (figure 2). With the donor-to-receiver distance of 16 μm and also taking dynamic effects such as overexpansion followed by oscillation of the rapidly expanded film into account, which can increase the blister height, the ns laser-induced blisters should contact the receiver and press the 2D material onto the Si/SiO$_2$ surface. The same could be expected for the fs laser pulses in the case of the short donor-to-receiver distance (figure 2). However, as mentioned above for ultrashort laser pulses, it is also probable that, due to the rapid deposition of laser energy to the titanium film at its interface with glass, titanium can swiftly vaporize in the interface region [53]. As a result, the Ti film, being inflated by the ablation products, will be expanding faster than in the case of purely thermal expansion and may involve plastic deformation of the film without it restoring to the initial state. Additionally, for the extremely fast laser heating of fs-irradiated films, the laser-produced stress is much higher compared to ns irradiation regimes [57,58]. This causes higher velocities of the film blistering process that also enables the ejection of all or nearly all deposited material within the blister area, overcoming the adhesion forces. Thus, we can conclude that, for fs laser pulses, the hBN flakes may be imprinted to the receiver by direct contact at the short donor-to-receiver distance while, for the larger distance, they are transferred via the gas phase with a success rate close to 100%.

To compare in detail the variation between near and far transfer, the 2D material on the donor and receiver films was analyzed for each blister. The areas approximating the blister spot on the donor surface were identified and the number of hBN flakes was counted for each blister. Similarly, all the material remaining outside the approximate blister area as well as that on the receiver was also counted. From the statistical analysis of the data sets from different pulse energies and donor-to-receiver distances (see Supplementary Material), it was found that, in the case of nanosecond laser pulses, more material is removed from the donor substrate at the 16-μm donor-to-receiver distance than for the 200-μm distance. For the fs-induced blisters, the difference between near and far transfer regimes is not statistically significant (see examples in figure 3 and Supplementary Material).

To transfer nanomaterial from a larger predefined area, experiments with successive, overlapped laser irradiation spots were performed (figure 4). In the case of ns irradiation, the majority of the hBN flakes can be transferred along the scan line (figure 4, A and B). However, with fs irradiation, blisters can be damaged and these can lead to some transfer of donor film material along with the hBN flakes (figure 4, C and D). The film damage/distortion is most likely caused by repeated deformation of the film at the overlapping areas under high laser-induced stresses inherent for fs laser pulses [58]. With each subsequent laser pulse, defect/damage accumulation takes place in the laser-affected film areas. From figure 4D, it can be seen that the first pulse (leftmost) leaves the irradiation spot almost unchanged, similar to those recorded for single pulses (figure 3, G and H). After the second pulse, a more distorted pattern is observed in the irradiated area while the subsequent 3$^{rd}$ and 4$^{th}$ pulses create cracks in the film, and titanium nanoparticles can be observed on the receiver substrate. The next pulses arrive on the partially damaged area. The presence of cracks reduces the stress around them [68], thus preventing efficient blistering, So, although some hBN flakes are transferred, there are also some 2D material flakes remaining on the irradiated regions of the donor, in contrast to what is observed with isolated laser pulses (figure 4, C and D).

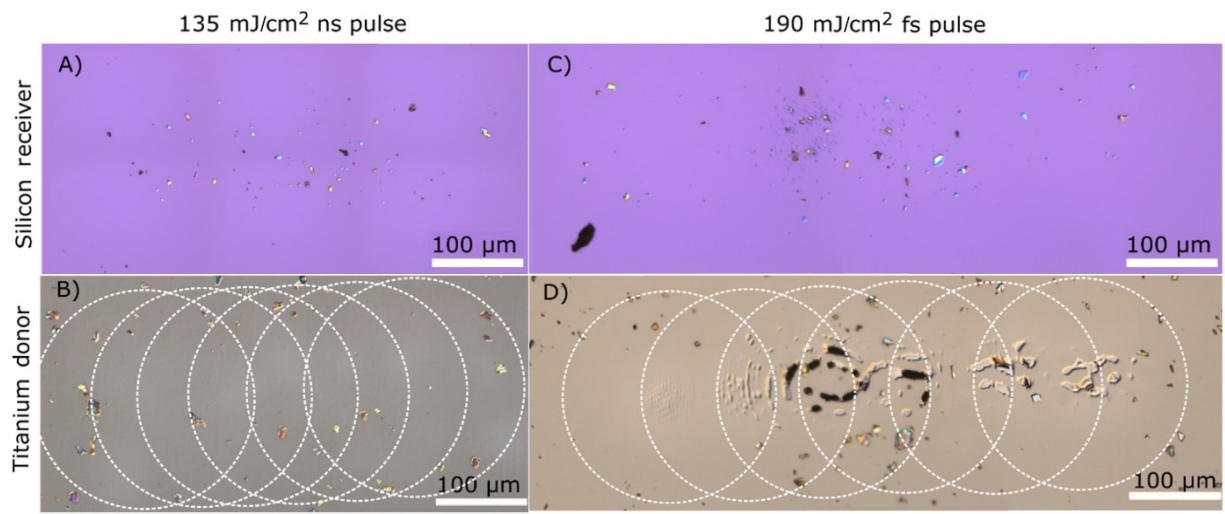

*Figure 4: BB-LIFT transfer in the laser scanning regimes with overlapping zones of irradiation. The direction of scanning is from left to right. A) and B) are the images for the ns laser irradiation regime at $F_0 = 135\pm15$ mJ/cm$^2$; C) and D) are obtained with fs laser irradiation at $190\pm10$ mJ/cm$^2$. A) and C) are the receiver substrates and B) and D) are the donors. These transfers for both pulse durations were at close proximity (16 μm). The areas of laser irradiation as determined by direct imaging of the beam are indicated on images B) and D) by dotted circles.*

In contrast to femtosecond pulses, the nanosecond laser-induced transfer enables efficient removal of material from the donor surface in experiments with overlapping pulses (figure 4, A and B). No cracks are seen in the film (figure 4B), removal is nearly complete in the central region of the scanning line, and 2D material is deposited on the receiver. For these experiments, the thermally expanded blisters can expand again and then re-contract toward the nearly flat initial film without any signs of transient melting. Thus, the film can be used as the donor substrate multiple times. From the results demonstrated above, it can be suggested that, under certain conditions of ns laser irradiation with close proximity receivers, laser-induced blister formation can enable stamping of nanomaterial to a receiver surface. Under optimal conditions, the blister expands to a large height that is followed by its collapse without any visible damage to the titanium film, thus enabling repeated laser exposure without the risk of transfer of DRL material. The velocity of the film expansion can reach several tens of ms$^{-1}$ [57] which is relatively small for a solid metal film to be destroyed upon stamping, thus avoiding the presence of any DRL fragments on the receiver.

The materials transferred to the surface at both donor-to-receiver distances appear to maintain their structural integrity as well as physical and chemical properties for both short and long donor-to-receiver distances. This has been verified previously for monolayer-thick semiconducting transition metal dichalcogenide flakes [45]. Figure 5 shows the Raman spectrum of a deposited multilayer hBN flake on the Ti DRL (A) and the Raman spectrum of a hBN flake on the receiver substrate (B), transferred using a ns pulse with $F_0 = 135$ mJ/cm$^2$. In figure 5(C), an optical microscope image of the transferred flake is given, showing a gradation in thickness, and hence number of layers, across the flake.

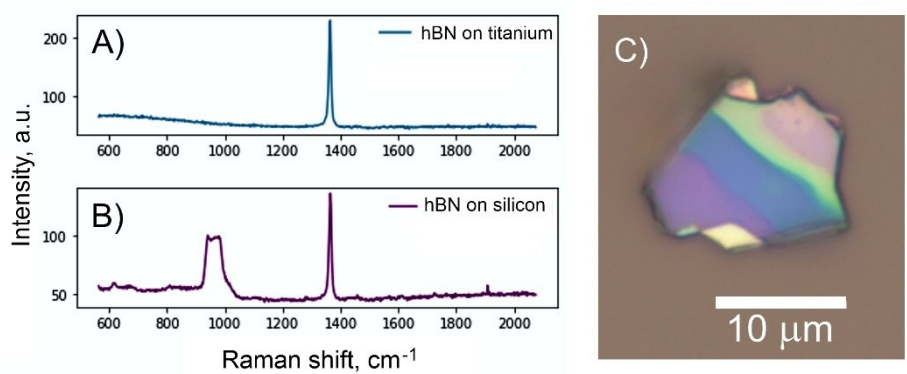

*Figure 5:* Raman spectra of hBN on the Ti donor surface (A) and on the SiO$_2$/Si receiver (B) transferred via ns BB LIFT at a donor-to-receiver distance of 200 μm at $F_0 = 135$ mJ/cm$^2$. The additional signal in (B) is due to the silicon underneath the hBN flake. The coloration seen in the optical microscope image of the transferred flake (C) is due to a variation in the number of layers across the flake.

Figure 6 shows an example of an AFM image of a monolayer hBN flake transferred onto an SiO$_2$/Si receiver. The thickness of the flake is near the limit of detectability. The height of the flake was measured as the difference between the average height over the whole flake (as defined more clearly by the phase map, figure 6(A)) and the average height of the surrounding region. As the hBN flakes were obtained via the "scotch-tape" method of exfoliation with subsequent cleaning via repeated soaking and washing with dichloromethane, there is still some residual adhesive as seen on the upper left of the flake. This adhesive bump was not included in the mask to measure the flake height as this would artificially increase it. According to this measurement, the thickness of the transferred hBN monolayer is ~0.3 nm, in agreement with the literature data [15]. From the combination of AFM, Raman, and optical imaging, it can be concluded that the nanomaterials transferred with BB LIFT do not show any signs of modification/degradation.

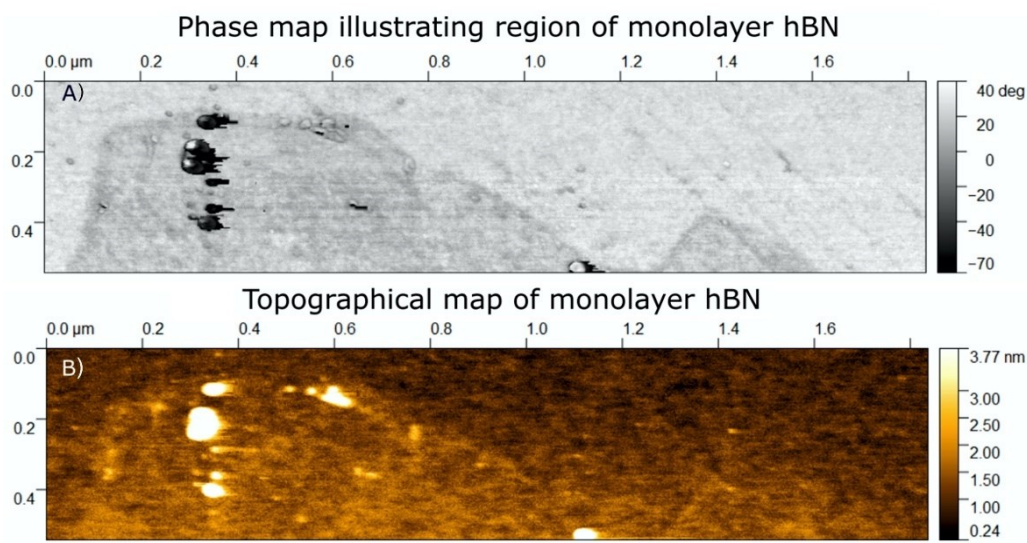

*Figure 6: AFM images of a monolayer hBN flake in the same region on the silicon receiver. A) phase map where the change in phase of tip resonance is measured. B) topographical image of this region where the hBN flake is much harder to discern.*

# Conclusions

This work demonstrates the effectiveness of the blister-based LIFT technique to transfer 2D flakes of hBN from a donor film with a 220 nm dynamic release layer of titanium deposited on a glass substrate. The transfer was performed under atmospheric conditions using femtosecond and nanosecond laser pulses for two donor-to-receiver distances of ~16 µm and 200 µm. From the data analysis and discussion of the mechanisms of blister formation, it was concluded that both irradiation regimes enable the "stamping" of the hBN flakes to the receiver surface for short donor-to-receiver distances (less than the estimated blister height) while, for larger distances, the transfer proceeds via mechanical ejection of nanomaterial. In the nanosecond irradiation regime studied here, the blister formation occurs via thermal expansion/bulging of the heat-affected area that enables the blister to cool and restore to the initial state after transferring the hBN flakes to the receiver surface. The blister height is affected by both the fluence, which increases the temperature of the film, and the radius of the blister. This gives a degree of freedom for a gentler low-energy transfer of material at low fluences via an enlarged laser beam radius. The ns low-energy BB-LIFT regime also allows a large-area transfer via scanning the beam with overlapped irradiation spots due to the absence of significant damage of the dynamic release layer by subsequent pulses. Femtosecond laser-induced blisters may be formed due to donor film ablation and the expansion of ablation products at the film-glass interface, additionally to the thermal expansion mechanism. However, this combined mechanism is more amenable to higher-resolution printing where the irradiation spot size is an important parameter. Both these mechanisms have been shown to be capable of transferring fragile nanomaterial without inducing damage. AFM images demonstrate the successful transfer of monolayer hBN flakes with an approximate thickness of 0.3 nm. Thus, it has been shown that the BB-LIFT enables a gentle and efficient transfer of 2D materials down to atomically thin layers. The findings of this research can be used for fabrication of micro/nanoelectronic and optoelectronic devices, sensors, and additive assembly of metamaterials and metasurfaces.

# Data availability statement

All data that support the findings of this study are included within the article and in Supplementary Information.

# Acknowledgements

I.M., A.V.B. and N.M.B. acknowledge support from the European Regional Development Fund and the State Budget of the Czech Republic (Project SENDISO No. CZ.02.01.01/00/22_008/0004596) and from the Ministry of Education, Youth and Sports of the Czech Republic through the e-INFRA CZ (ID:90254). N.M.B. also acknowledges support from COST Action CA23125 TETRA "The mETamaterial foRmalism approach to recognize cAncer" (European Cooperation in Science and Technology). The authors are grateful to T. Hotta and R. Kitaura for stimulating discussions.

# ORCID iDs

N T Goodfriend https://orcid.org/0000-0003-1195-0901
I Mirza https://orcid.org/0000-0003-4450-9901
A V Bulgakov https://orcid.org/0000-0002-9651-1328
E E B Campbell https://orcid.org/0000-0002-4656-3218
N M Bulgakova https://orcid.org/0000-0003-1449-8535

# Supplementary Material

## Laser-Induced Gas-Phase Transfer and Direct Stamping of Nanomaterials: Comparison of Nanosecond and Femtosecond Pulses

Nathan T. Goodfriend[1], Inam Mirza[1], Alexander V. Bulgakov[1], Eleanor E.B. Campbell[2,3], Nadezhda M. Bulgakova[1]

[1] FZU – Institute of Physics of the Czech Academy of Sciences, Na Slovance 1999/2, 182 00 Prague, Czechia

[2] EastCHEM, School of Chemistry, University of Edinburgh, David Brewster Road, Edinburgh EH9 3FJ, United Kingdom

[3] Dept. of Physics, Ewha Womans University, Seoul 03760, Republic of Korea

### 1. *Transfer Efficiency: Data Analysis*

Multiple sets of 2D hBN were transferred from a donor film to a receiver surface at different laser pulse energies, different donor-to-receiver distances, and different laser pulse durations. The transfer was performed from different locations on the donor substrate. Due to the nature of the deposition on the dynamic release layer (DRL), there is a large spread in the number of particles in any different irradiated spot area with an average hBN flake density of $860 \pm 490$ $\mu m^{-2}$ across the DRL. The hypothesis presented in the main manuscript is that when the donor is located in close proximity to the receiver, the blister may make physical contact with the receiver during expansion when irradiated with laser pulses. This can impact the efficiency of material transfer. From a consideration of the images in Fig 3 of the main text it appears that the transfer when using ns pulses is more efficient for the close distance (16 μm) compared to the far distance (200 μm). For fs pulses a high transfer efficiency is indicated for both separations. To identify if there is a statistically significant difference in the transfer efficiency for the different conditions, a detailed investigation of the number of particles remaining on the blister area and transferred to the receiver was carried out.

The number of hBN flakes remaining within the blister area and the number transferred to the receiver were recorded for all laser conditions and both transfer distances. The blister area for ns irradiation, corresponding to the area of the laser spot on the DRL was $1.9 \times 10^4$ $\mu m^2$ while for fs irradiation it was $1.5 \times 10^4$ $\mu m^2$. The area did not change significantly with changing pulse fluence in both cases. To compare the different sets of data, the average transfer probability was calculated for each data set, assuming that no particles were "lost" on transfer, thus avoiding the problem of the large variation in particle density on the donor substrate. The average transfer probabilities for each laser fluence and separation distance are tabulated in Tables 1 and 2 for ns BB-LIFT and fs BB-LIFT, respectively. The students' two-tail T-test [1] was used to determine if there was a statistically significant variation in the mean transfer probability between the sets. In each case the F-test [1] was first applied to determine whether there was a significant difference in the variances of any two data sets before applying the relevant two-tail T-test (equal or non-equal variances). The null hypothesis for the tests is that the transfer probability is not significantly different for the compared data sets. An entrance of "TRUE" in the following Tables 4 and 5 indicates that there is no statistically relevant difference between the average transfer probabilities. Data sets were only considered where there was no significant observable disruption of the donor layer.

**Table 1**. Summary of results for ns BB-LIFT

| Fluence/ mJ/cm$^2$ | Distance/ μm | av. transfer probability /% | Standard Deviation | Number of Measurements |
|---|---|---|---|---|
| 135 | 16 | 83.4 | 15.0 | 10 |
| 135 | 200 | 39.0 | 10.5 | 9 |
| 170 | 16 | 90.4 | 5.4 | 10 |
| 170 | 200 | 57.3 | 18.3 | 7 |
| 220 | 16 | 79.7 | 13.7 | 9 |
| 220 | 200 | 70.0 | 8.7 | 9 |
| 290 | 16 | 86.5 | 12.0 | 8 |
| 260 | 200 | 88.3 | 3.3 | 8 |

**Table 2**. Summary of results for fs BB-LIFT

| Fluence/ mJ/cm$^2$ | Distance/ μm | av. transfer probability /% | Standard Deviation | Number of Measurements |
|---|---|---|---|---|
| 180 | 16 | 93 | 2.9 | 6 |
| 190 | 16 | 86 | 4.7 | 6 |
| 198 | 16 | 87 | 13.5 | 6 |
| 203 | 200 | 82 | 10.9 | 7 |
| 217 | 200 | 75 | 5.7 | 3 |
| 227 | 200 | 77 | 17.6 | 3 |

## 2. *ns BB-LIFT*

The results of applying the two-tail T-test for different laser fluences used for ns BB-LIFT are summarized in Tables 3-5. Table 3 gives the results of testing the average transfer probability of hBN flakes for different laser fluences when there is a short (16 μm) distance between the donor film and receiver substrate. An entrance of "TRUE" in the table indicates that there is no statistically relevant difference. This supports the hypothesis of efficient transfer happening via "stamping of the material on the receiver by the expanding blister, as long as the expanded blister reaches the receiver. The overall average transfer probability is 85 ± 13%. For all investigated fluences, the blister is estimated to expand to a height greater than the separation (see Fig. 2 of the main manuscript).

In contrast, the comparison of the transfer probability for the larger donor-receiver distance using ns lasers shown in Table 4 does depend on the fluence. The transfer probability generally increases as the laser fluence increases as can be seen in Table 1.

**Table 3**. Comparison of the average transfer probability for hBN flakes with a small 16 μm separation between donor and receiver substrates for different ns laser fluences.

| mJcm$^{-2}$ | 135 | 170 | 220 | 290 |
|---|---|---|---|---|
| 135 | - | TRUE | TRUE | TRUE |
| 170 | TRUE | - | TRUE | TRUE |
| 220 | TRUE | TRUE | - | TRUE |
| 290 | TRUE | TRUE | TRUE | - |

**Table 4**. Comparison of the average transfer probability for hBN flakes with a large, 200 μm, separation between donor and receiver substrates for different ns laser fluences. Plus and minus signs for a FALSE outcome indicate when the difference between data set 1 (column) and data set 2 (row) is positive or negative.

| mJcm$^{-2}$ | 135 | 170 | 220 | 260 |
|---|---|---|---|---|
| 135 | - | FALSE – | FALSE – | FALSE – |
| 170 | FALSE + | - | TRUE | FALSE – |
| 220 | FALSE + | TRUE | - | FALSE – |
| 260 | FALSE + | FALSE + | FALSE + | - |

Table 5 compares the ns transfer probabilities for short (S) and long (L) donor-receiver distances. With the exception of the largest fluence investigated at a separation of 200 μm, the results clearly show that transfer over the short distance is significantly more successful than that over the longer distance. As the laser fluence increases, the average transfer probability for the larger distance increases until it is no longer significantly different from the short distance transfer. However, for such high laser fluences there is a higher probability for donor substrate disruption and transfer of metal to the receiver.

**Table 5**. Comparison of average transfer probability for particles transferred over short (S, 16 μm) and long (L, 200 μm) distances for ns BB-LIFT

| mJcm$^{-2}$ | 135 L | 170 L | 220 L | 260 L |
|---|---|---|---|---|
| 135 S | FALSE + | FALSE + | FALSE + | TRUE |
| 170 S | FALSE + | FALSE + | FALSE + | TRUE |
| 220 S | FALSE + | FALSE + | TRUE | TRUE |
| 260 S | FALSE + | FALSE + | FALSE + | TRUE |

Figures 1 and 2 show the donor film surface after ns BB-LIFT with a fluence of 135 mJcm$^{-2}$. for the long and short separations, respectively. The difference is very striking with the short distance when the donor areas are almost clear whereas significantly more particles remain on the long distance examples.

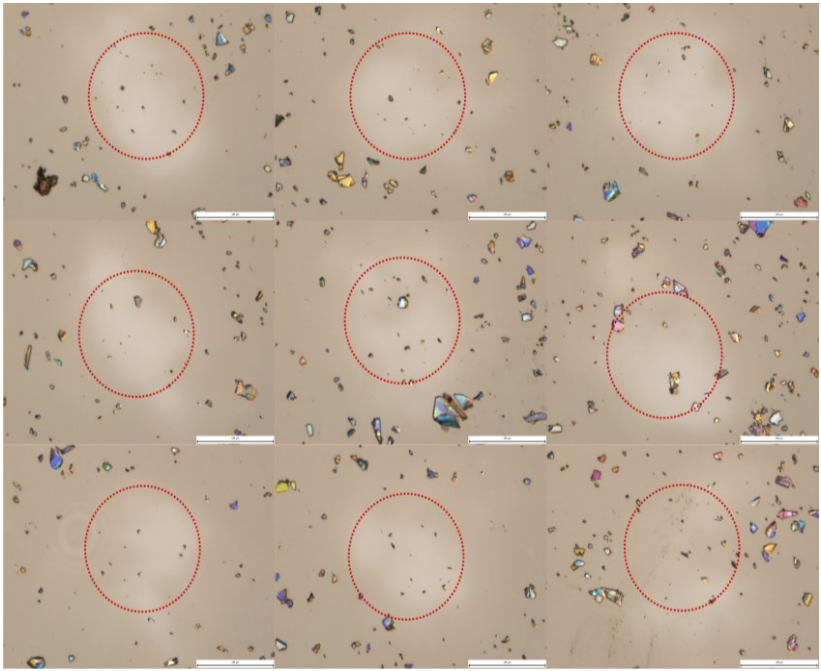

**Fig. 1** Donor areas after ns transfer at 135 mJ/cm$^2$ to a receiver at a distance of 200 μm.

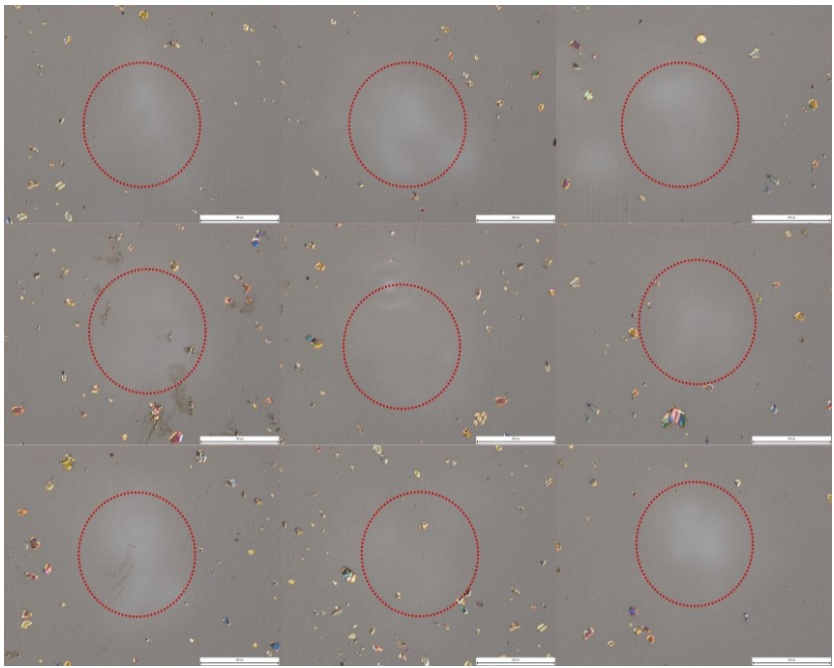

**Fig. 2** Donor areas after ns transfer at 135 mJ/cm$^2$ to a receiver at a distance of 16 μm.

### 3. *fs BB-LIFT*

A similar analysis was carried out for the fs laser pulses. Table 6 summarizes the comparison between different laser fluences for the 16 μm separation and Table 7 for the 200 μm separation.

**Table 6**. Comparison of the average transfer probability for hBN flakes with a small 16 μm separation between the donor and receiver substrates for different fs laser fluences.

| mJcm$^{-2}$ | 180 | 190 | 198 |
|---|---|---|---|
| 180 | - | FALSE + | TRUE |
| 190 | FALSE - | - | TRUE |
| 198 | TRUE | TRUE | - |

**Table 7**. Comparison of the average transfer probability for hBN flakes with a large 200 μm separation between the donor and receiver substrates for different fs laser fluences.

| mJcm$^{-2}$ | 203 | 217 | 227 |
|---|---|---|---|
| 203 | - | TRUE | TRUE |
| 217 | TRUE | - | TRUE |
| 227 | TRUE | TRUE | - |

All fs data sets can be considered to belong to the same distribution with the exception of the lowest fluence transfer with a short distance. This case has a significantly higher transfer probability than the other short distance measurements. It is not clear why this should be and it may simply be a consequence of the larger number of data points and lower standard deviation compared to the higher fluences. In contrast to the ns transfer, there is no significant fluence dependence of the transfer probability for the 200 μm separation. This could be providing evidence for the predominance of the ablative mechanism for fs pulses but the range of fluences is too small to provide a conclusive argument.